# Machine learning for decision-making under uncertainty

Lizhi Xin[1], Kevin Xin[2], Houwen Xin[3, *]

[1]Building 59, 96 Jinzhai Road, Hefei, Anhui, P. R. China
[2]6 South Laflin Street Chicago, IL, USA
[3]Department of Chemical physics USTC, Hefei, Anhui, P. R. China
[*]hxin@ustc.edu.cn

## ABSTRACT

We live in a world brimming with uncertainty, where we constantly have to make a lot of decisions under incomplete information. We are firm believers that our subjective belief cannot be computed by rigorous mathematical formula; instead based on Darwin's natural selection (the evolution process is simulated by machine learning with genetic programming), a proposed computational model that incorporates insights from quantum theory to describe and explain decision-making under uncertainty. Unlike other decision-making theories that explain the decision-making process through probability theory, our proposed decision theory discovers "laws" of thought by learning observed historical data. There is no differential equation and no transition probability in our decision theory, our decision model has an emphasis on machine learning, where decision-makers build-up their experience by being rewarded or punished for each decision they make, and prepare them for making better decisions in the future. We do not model with the usual utility function, but with quantum decision tree that simulates people's decision process. Each quantum decision tree includes a set of strategies; every time a decision is made, the decision-maker first chooses a strategy from the quantum decision tree's strategy pool, and then chooses an action based on the degree of belief which is obtained by genetic programming based on maximizing expected value.

## Introduction

How we make decisions is truly an enigma. Gerolamo Cardano laid down the fundamentals of the then-new field of probability – a new way to describe the chance of how likely something was to happen[1]. Blaise Pascal and Pierre de Fermat discussed the gambling question asked by Chevalier de Méré and proposed expected value theory to make decisions, which is to select the biggest expected value out of the all-possible selections. Daniel Bernoulli proposed the expected utility decision theory to solve the St. Petersburg Paradox (which went against expected value theory). John von Neumann and Oscar Morgenstern were the first to axiomatize expected utility theory[2]. Leonard "Jimmie" Savage put forth his theory of subjective utility[3]. The Allais Paradox and the Ellsberg Paradox[4-5] were developed to refute the von Neumann-Morgenstern objective expected utility theorem and Savage's subjective utility theory, respectively. Both of these paradoxes have showed that when we are faced with a choice of attaining something with more certainty over something that has a considerable amount of risk though with a big reward, we tend to decide to take the former. Because of our tendencies to be more risk aversive when faced with choosing a small-probabilistic event from happening, Daniel Kahneman and Amos Tversky, developed prospect theory[6] that has detailed how individuals assess the chance of losing and gaining in a relative asymmetric manner. However, the human behaviors exhibited during decision-making, such as the order effect, cannot be sufficiently explained by decision theory based on classical probability.

Classical decision theory is a "black box". Scientists are trying to apply quantum theory to reveal how decisions are made. Recently many quantum-like decision theories[7-9] have been proposed based on quantum probability to revise the mathematical structure that's used in classical models. Aerts et al. first proposed to apply quantum probability in decision theories[10-11]; Busemeyer et al. proposed a quantum-like model to describe human judgments and the order effect[12-14]; Khrennikov et al. improved the Busemeyer quantum model by applying quantum instrument of quantum measurement theory[15-19]; Yukalov et al. proposed a rigorously axiomatic quantum decision theory[20-22].

Whether it is classical decision theories or quantum-like decision theories, all well-developed decision models have applied a rigorous mathematical structure to describe people's decision-making under uncertainty. The main issue with mathematical models is that they are difficult to understand, cannot reflect the dynamic changes in the state of the decision-makers' mind, and is not easy to calculate theoretical values to compare with actual observed outcomes when the mathematical model becomes more complex.

In 2019, Susan Athey and Guido Imbens quoted Leo Breiman in their paper "Machine Learning Methods That Economists Should Know About[23]":

"There are two cultures in the use of statistical modeling to reach conclusions from data. One assumes that the data are generated by a given stochastic data model. The other uses algorithmic models and treats the data mechanism as unknown… Algorithmic modeling, both in theory and practice, has developed rapidly in fields outside statistics. It can be used both on large complex data sets and as a more accurate and informative alternative to data modeling on smaller data sets. If our goal as a field is to use data to solve problems, then we need to move away from exclusive dependence on data models and adopt a more diverse set of tools."

In this paper, based on Darwin's natural selection (the evolution process is simulated by machine learning with genetic programming), we propose a computational model (not a rigorous mathematical model) that incorporates insights from quantum theory to describe and explain people's decision-making under uncertainty. Our decision model emphasizes machine learning, where decision-makers build-up their experience by being rewarded or punished for each decision they make, and prepare them for making better decisions in the future. This is more in line with decision-makers in the real world.

Basically, our proposed quantum decision theory discovers "laws" of thought by just machine-learning observed historical data (sequential measurement of observables); there is no differential equation, and no transition probability computation in our decision theory. We do not model with the usual utility function or observables of the projection type in other quantum-like decision theories, but with quantum decision tree. The quantum decision tree, which simulates people's decision process, can be interpreted as a mixed strategy. Each quantum decision tree includes a set of strategies, each of which is a mixed density operator; every time a decision is made, the decision-maker first chooses a strategy from the quantum decision tree's strategy pool, and then chooses an action based on the degree of belief; the decision maker's subjective probability is obtained by genetic programming based on maximizing expected value. Thus, subjective probability is obtained through Darwin's natural selection not computed by classical or quantum probability theory. Our algorithmic model assumes decision-makers' subjective probability is unknown and can only be approximately obtained through machine learning the historical data sequence.

**Decision-making process**

Anytime when we make a decision, there are the natural states of the world that we have to take into account, and then there are actions we can choose to take. The natural states of the world are objective, and our beliefs are subjective, which are influenced by nature and ultimately affect what actions we take.

Decision-making under uncertainty can be represented by Table 1. We subjectively choose an action $a_i \in \{a_1, a_2, \ldots, a_n\}$ where nature's objective state is in $q_i \in \{q_1, q_2, \ldots, q_n\}$ when a decision is made, and the result of the decision (value matrix $v_{ij}$) depends on both the state of nature and choice of the action taken.

| State \ Action | $q_1$ | … | $q_j$ | … | $q_n$ |
|---|---|---|---|---|---|
| $a_1$ | $v_{11}$ | | … | | $v_{1n}$ |
| ⋮ | | ⋱ | | | |
| $a_i$ | ⋮ | | $v_{ij}$ | | ⋮ |
| ⋮ | | | | ⋱ | |
| $a_m$ | $v_{m1}$ | | … | | $v_{mn}$ |

**Table 1** State-action-value decision table

Using the futures market as an example, Table 2 shows the two states of $q_1$ and $q_2$ signify that the market is rising and falling, respectively. Then we have the two possible actions, $a_1$ and $a_2$ being buy and sell, respectively. All together, we have all the possible outcomes of $v_{11}$, $v_{12}$, $v_{21}$, and $v_{22}$, which signify whether we profited or not.

| State / Action | $q_1$ | $q_2$ |
|---|---|---|
| $a_1$ | $v_{11}$ | $v_{12}$ |
| $a_2$ | $v_{21}$ | $v_{22}$ |

**Table 2** State-action-value decision table of futures market

The market influences the traders' decisions, while the traders' actions then decide the market's state. This interaction between the two, the objective (state of the market) and the subjective (traders' beliefs) is what causes both the result of the decisions (gain or loss) and the state of the market (up or down) to be uncertain.

The state of the market describes the objective world; it can be represented by the superposition of all possible states in terms of the Hilbert state space as shown below[24-25].

$$|\psi\rangle_{\text{market}} = c_1|q_1\rangle + c_2|q_2\rangle \qquad |c_1|^2 + |c_2|^2 = 1 \tag{1}$$

Where $|q_1\rangle$ denotes a state in which the market has gone up and $|q_2\rangle$ denotes a state in which the market has gone down. $|c_1|^2$ is the objective frequency of the rising; $|c_2|^2$ is the objective frequency of the falling market.

The state of the trader's mind is the subjective world. We postulate that when the trader is undecided in making a trade (buy or sell), it can be represented by superposition of all possible actions as follows.

$$|\phi\rangle_{\text{mind}} = \mu_1|a_1\rangle + \mu_2|a_2\rangle \qquad |\mu_1|^2 + |\mu_2|^2 = 1 \tag{2}$$

Where $|a_1\rangle$ denotes the trader's action to buy, and $|a_2\rangle$ denotes the trader's action to sell. $|\mu_1|^2$ is the trader's degrees of belief in betting that the market will rise; $|\mu_2|^2$ is trader's degrees of belief in betting that the market will fall.

The external objective world, the market, and the subjective world, the trader with their beliefs, all come together into action whenever a trader goes to trade. The information that they have available to them prior making their decision is limited, in a sense, basically incomplete, which then force traders to essentially "guess" or "bet". They don't know whether the market will rise or fall. The value of the trader's decision (profit or loss) is uncertain; we hypothesize that it can be represented by a mixed state's density operator as a value operator in (3). In quantum mechanics, all observable variables are expressed by operators. The value is expressed by an operator because value is an observable variable. Classical decision theory uses value (utility) functions to evaluate people's decisions (gains or losses). In our proposed quantum decision theory, we use value operator to evaluate people's decisions.

$$\hat{V} = p_1|a_1\rangle\langle a_1| + p_2|a_2\rangle\langle a_2| \qquad p_1 + p_2 = 1 \tag{3}$$

Where $p_1$ is the subjective probability of choosing $a_1$ (buying), and $p_2$ is the subjective probability of choosing $a_2$ (selling).

Before a trader makes a decision, his/her mind state is in a pure state, a superposed state in which they can decide whether to buy and sell at the same time. But in reality, you can't take an action to buy and sell simultaneously. This pure state is when the states of buy and sell are superposed in the trader's mind. Then when the traders' make the decision, the state of the traders' mind is then transformed from that pure state into a mixed state, which is when they decide to buy or sell, with certain degrees of belief. Basically, this transformation is the trader choosing from one of the available actions, with action $a_1$ being buy with probability $p_1$ and action $a_2$ being sell with probability $p_2$, shown below.

$$D: |\phi\rangle\langle\phi| \xrightarrow{\text{decision}} \hat{V} = p_1|a_1\rangle\langle a_1| + p_2|a_2\rangle\langle a_2| \tag{4}$$

Information is the essence of people's subjective beliefs just like energy is the essence of the objective world. Claude Shannon said, "Information is the resolution of uncertainty," we believe that it is not just information that resolves uncertainty, but valuable information is what reduces it. Though the subjective will always be influenced by the objective to certain degree, it is the unification of the two that will bring along much needed valuable information when making a decision. And it is here that we now take the two and unify them into one, as quantum expected value(qEV), as in (5).

$$\begin{aligned} qEV &= \langle\psi|\hat{V}|\psi\rangle = (c_1^*\langle q_1| + c_2^*\langle q_2|)(p_1|a_1\rangle\langle a_1| + p_2|a_2\rangle\langle a_2|)(c_1|q_1\rangle + c_2|q_2\rangle) \\ &= p_1\omega_1|\langle a_1|q_1\rangle|^2 + p_1\omega_2|\langle a_1|q_2\rangle|^2 + p_2\omega_1|\langle a_2|q_1\rangle|^2 + p_2\omega_2|\langle a_2|q_2\rangle|^2 \\ &= p_1\omega_1 v_{11} + p_1\omega_2 v_{12} + p_2\omega_1 v_{21} + p_2\omega_2 v_{22} \\ &= \sum_{i=1,2} p_i \sum_{j=1,2} \omega_j v_{ij} \end{aligned} \quad (5)$$

Where $p_i$ is a trader's subjective probability in choosing an action $a_i$, subjective probability represents the trader's degrees of belief in a single trade; $\omega_j$ is the objective frequency at which state of the futures market is in $q_j$, objective frequency represents the statistical results of multiple occurrences of objective states; matrix $v_{ij}$ is the value when the trade was made, in which the trader chose an action $a_i$ where state is in $q_j$ as shown below.

$$v_{ij} = |\langle a_i|q_j\rangle|^2 = \begin{cases} 1, & i = j \\ -1, & i \neq j \end{cases} \quad (6)$$

When the trader "guesses" the direction of the market correctly (i = j), then he/she profits. This could be if the market goes up and they buy or if the market goes down and they sell. If the trader doesn't "guess" correctly (i ≠ j), then he/she suffer a loss. This would be if the market goes up but the trader sells, or if the market goes down and the trader buys. The different actions that the trader's took lead to different value (gain or loss); in other words, the value is "created" based on both traders' subjective beliefs and objective natural states.

The quantum expected value (qEV) can be seen in two distinct parts – an "average of average" like situation. This is simply taking the subjective and the objective then assigning them to two respective parts of the full qEV, resulting in the unification of the two.

$$oEV_i = \sum_{j=1,2} \omega_j v_{ij} \quad (7)$$

We have the first average, the objective expected value, which we will term as in equation (7). This is the part of the objective world, the external info that we take in.

$$sEV_i = p_i oEV_i \quad (8)$$

Next, we have the second part, which is the subjective expected value as in equation (8). In order to get the average of average, the unified qEV, we take the two, the $oEV_i$ and the $sEV_i$, and then we average it, or mathematically, $qEV = \sum_{i=1,2} sEV_i$. All together this becomes the subjective expected value of the objective expected value (average of average), which is the unification of the objective natural world and the subjective human beliefs.

From the great classical theories of von Neumann and Savage, their theories are the two extremes: von Neumann is all objective, while Savage is all subjective. Ours is the unified both objective and subjective.

## Machine learning for decision-making under incomplete information

Basically the quantum value operator is just a 2x2 matrix which it can be constructed approximately with the 8 basic quantum gates[26] and 3 logic operations as a quantum decision tree (qDT). A qDT simulates the decision-making process. The qDT will guide people to choose corresponding actions based on their subjective beliefs through the objective world. A qDT will have branches and nodes, which will be further categorized as leaf nodes and non-leaf nodes. Non-leaf nodes are represented by operation set F, and leaf nodes are represented by data set T, both shown in (9) and (10).

$$F = \{+(ADD), \quad *(MULTIPLY), \quad //(OR)\} \tag{9}$$

$$T = \{H, X, Y, Z, S, D, T, I\} \tag{10a}$$

$$\left\{ \begin{array}{l} H = \frac{1}{\sqrt{2}}\begin{bmatrix} 1 & 1 \\ 1 & -1 \end{bmatrix}, X = \begin{bmatrix} 0 & 1 \\ 1 & 0 \end{bmatrix}, Y = \begin{bmatrix} 0 & -i \\ i & 0 \end{bmatrix}, Z = \begin{bmatrix} 1 & 0 \\ 0 & -1 \end{bmatrix} \\ S = \begin{bmatrix} 1 & 0 \\ 0 & i \end{bmatrix}, D = \begin{bmatrix} 0 & 1 \\ -1 & 0 \end{bmatrix}, T = \begin{bmatrix} 1 & 0 \\ 0 & e^{i\pi/4} \end{bmatrix}, I = \begin{bmatrix} 1 & 0 \\ 0 & 1 \end{bmatrix} \end{array} \right\} \tag{10b}$$

The actual construction process of a qDT is by randomly selecting a logic symbol from the operation set F, the non-leaf nodes, which are the "branches" that grow corresponding accordingly to the nature of the operation symbol and so on until a leaf node is reached. A fully grown qDT has the non-leaf nodes as circles and the leaf nodes as squares.

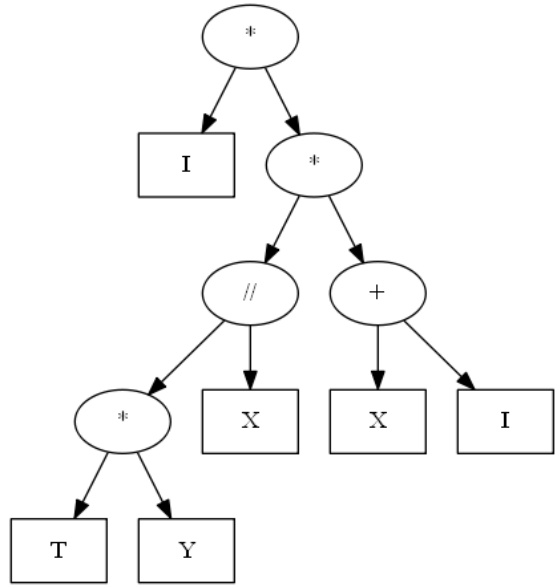

**Figure 2** A sample qDT.

Figure 2 is a sample qDT. We start off with the tree's "root", in this case is the multiplication operation in the top circle. By rules of multiplication, it "grows" two branches, the "I" quantum gate in the square and another multiplication operated in a circle. Since the quantum gate "I" is a leaf node, it ends there. Since the second multiplication is a non-leaf node, meaning it will keep growing from that one. It keeps growing, resulting in two more operation symbols, an OR (//) and an ADD (+). From here, they both go onto branch off to more nodes. The OR grows another multiplication operator and an "X" quantum gate leaf node. The addition grows two leaf nodes, therefore ending it with an "X" quantum gate and an "I" quantum gate. The multiplication that arose from the OR, then continues to grow branches, while the "X" quantum gate produced by the OR ends. And finally, we get to the two "leaves" produced by the * from the //, a "T" and "Y" quantum gates, completing this qDT. Mathematically this qDT is expressed as in (12):

$$qDT = \left( I * \left( ((T * Y)//X) * (X + I) \right) \right) \tag{12a}$$

- $S_1 = \left( I * (X * (X + I)) \right) \rightarrow \hat{V} = |a_1\rangle\langle a_1| \tag{12b}$

- $S_2 = \left(I * ((T * Y) * (X + I))\right) \rightarrow \hat{V} = 0.04|a_1\rangle\langle a_1| + 0.96|a_2\rangle\langle a_2|$ (12c)

According to this qDT, what it tells us is that there are two different strategies that we can take: strategy $S_1$ is 100% certain to take action $a_1$ (buy); and strategy $S_2$ is 96% degree of belief to take action $a_2$ (sell) and a 4% degree of belief to take action $a_1$ (buy). This shows that the two strategies respective subjective probabilities of this qDT are almost completely close to unity, resonating almost maximum information. Just a quick elaboration on why this qDT produces two strategies. Since there is an OR operation symbol, that means that the two branches under the OR have a 50/50 chance of being used, one of which is the "X" gate and the other is the (T * Y). Thus, that's why are two different strategies produced by qDT.

In the words of Ray Solomonoff in his paper "*Does Algorithmic Probability Solve the Problem of Induction?*", "We are uncertain of the probabilities needed for the decision and we cannot express this uncertainty probabilistically." Hence, this is the exact reason why we use qDT, because we first select one strategy from a group of strategies provided by the qDT, and then select one action with subjective probability to deal with the uncertainty need of the decision. The nested hierarchy structure of qDT attempts to solve the problem of "we cannot express this uncertainty probabilistically."

qDT alone will not aid us into making the best decisions, the next step is to find a way to optimize qDT with a group of satisfactory strategies to guide our decisions. To optimize anything, there needs to be: first, a selection of a good evaluation function and two, how to acquire an optimal solution. First off, in our model, any decision-maker will try to maximize their qEV (quantum expected value) when making any decision. Thus, we need to evaluate how "fit" the result (profit or deficit) of the trader's decision, which we can do so by using qEV as a fitness function to optimize the qDT by evolving them (using qGP).

What the fitness function essentially is a particular type of function that is used to summarize, as a single figure of merit, how close a given design solution is to achieving the set aims. Fitness functions are used in genetic programming[27-28] to guide simulations towards optimal design solutions. In order to reach the optimal solution, the qGP algorithm implements a continuous evolution process through selection, crossover, and mutation. The whole idea of having qGP go through an iterative evolution loop is to find a satisfactory qDT by means of learning historical data to obtain the most optimal solution.

The learning rules are as follows:

a) If the futures market is up ($q_1$)

    i. If the trader bets the market is up ($a_1$), the trader profits ($v_{11} = 1$);

    ii. If the trader bets the market is down ($a_2$), the trader deficits ($v_{21} = -1$);.

b) If the market is down ($q_2$)

    i. If the trader bets the market is down ($a_2$), the trader profits ($v_{22} = 1$);

    ii. If the trader bets the market is up ($a_1$), the trader deficits ($v_{12} = -1$);.

The result ($qEV_k$) of a kth trade as follows:

$$qEV_k = \begin{cases} p_1\omega_1, & |\phi\rangle_{mental} = |a_1\rangle \text{ and } |\psi\rangle_{market} = |q_1\rangle \\ -p_1\omega_2, & |\phi\rangle_{mental} = |a_1\rangle \text{ and } |\psi\rangle_{market} = |q_2\rangle \\ -p_2\omega_1, & |\phi\rangle_{mental} = |a_2\rangle \text{ and } |\psi\rangle_{market} = |q_1\rangle \\ p_2\omega_2, & |\phi\rangle_{mental} = |a_2\rangle \text{ and } |\psi\rangle_{market} = |q_2\rangle \end{cases} \quad (13)$$

It is critical to note here that from the above four possible outcomes only one can actually happen. Because the market can only be up or down, and a trader's choices are only to buy or sell, only one of those combinations of whether the market is up or down and if the trader will buy or sell can occur in reality. $\omega_1$ and $\omega_2$ (the frequencies of whether the market is up or down) are known variables, which are approximately 50/50. Now, there are a couple of things that we don't know. One, what action a trader will take at any given time, since everyone is different. Two, we also don't know how "confident" a trader is in their choice when they take their action, which is either buy or sell based off of the info and beliefs they have. This is represented by $p_1$ and $p_2$, which are their beliefs that the market will go up or down, respectively. The whole point of optimizing all of this is for two reasons: one, getting the trader to make a decision as correct as possible, and two, to do so with the highest degree of beliefs, which would mean either $p_1$ or $p_2$ is equal to 1. Essentially this is what qDT is trying to do: get the trader to make the best decision every single time, and by doing so the chosen strategy needs to be "selected" correctly every time with the greatest degree of belief. The fitness function represents the sum of all the qEV$_k$ by learning all the historical data as shown below.

$$f_{fitness} = \sum_{k=0}^{N} qEV_k \quad (14)$$

The optimal solution is implemented by a qGP algorithm, below.

**qGP algorithm**
*Input*:
- Historical data set $\{d_k = (q_k, x_k), k = 0, \cdots, N\}$.
- Setting
1) Operation set $F = \{+, *, //\}$
2) Data set $T = \{H, X, Y, Z, S, D, T, I\}$, eight basic quantum gates
3) Crossover probability = 75%; Mutation probability = 5%.

*Initialization*:
- Population: randomly create 300 qDTs.

*Evolution*:
- for $i = 0$ to n (n=100 generations)
a) Calculate fitness for each qDT based on historical data set.
b) According to the quality of fitness:
   i. Selection: selecting parent qDTs.
   ii. Crossover: generate a new offspring using the roulette algorithm based on crossover probability.
   iii. Mutation: randomly modify parent qDT based on mutation probability.

*Output*:
- A qDT of the best fitness.

There are four parts that are in sequential order when the algorithm is applied: Input, Initialization, Evolution, and Output. In the input phase, historical data is gathered to learn from ($q_k$ denotes state of equity and $x_k$ denotes closing price). Then the qDT that will be constructed by the algorithm are "set", with the operation and the quantum gates. For the crossover and mutation probabilities, what that basically means is that the qDT have a 75% chance of producing an offspring, or child, and a 5% of having one of its own branches modified. Then the initialization phase, 300 qDTs are randomly created as population. The evolution stage is pretty much self-explanatory; it's basically a survival of the fittest of the generated qDTs, with the strongest surviving and passing down their "genes" to their "children". Output is then the best qDT that have evolved.

## Results

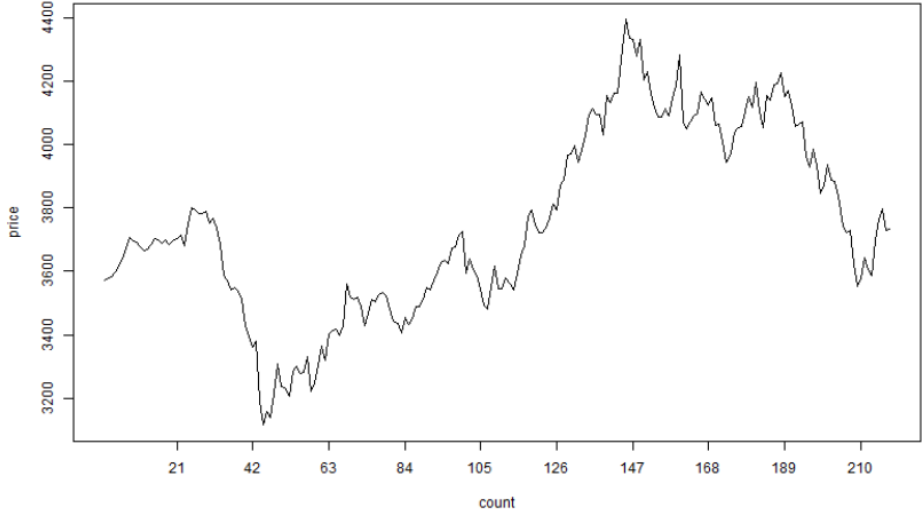

**Figure 3** Price fluctuation of rebar (contract rb1901) traded on Shanghai Futures Exchange.

The data of rebar rb1901 (transaction cycle is day from 2018/1/16 to 2018/12/7) traded on the Shanghai Futures Exchange is used as the historical data for optimization. It is the 218 closing prices throughout the duration of the contract. Just like any human would go about entering the world of trading, most wouldn't go into trading blind, it would probably be best if they look at past historical trading data to formulate their trading strategies. By following the learning rules, the qGP algorithm will use this historical data as training data to search for satisfactory qDTs by maximizing qEV. Below is one satisfying qDT (Figure 4) that includes a group of strategies that's been evolved by qGP to guide a trader to take actions which allows him/her to make the greatest possible profit:

$$qDT = \left( \left( D * \left( Y + \left( \left( \left( \left( X // (D * (S + Z)) \right) // I \right) // I \right) // (T + H) \right) + T \right) \right) \right) // \left( ((I * Y) + (T // Y)) + (I // Y) \right) \right) \tag{15}$$

This qDT can be broken down by the 8 strategies it produces:

- $S_1 = \left( ((I * Y) + T) + I \right) \rightarrow \hat{V} = 0.88|a_1\rangle\langle a_1| + 0.12|a_2\rangle\langle a_2|$ (88% belief to buy, 12% belief to sell)
- $S_2 = \left( ((I * Y) + T) + Y \right) \rightarrow \hat{V} = 0.83|a_1\rangle\langle a_1| + 0.17|a_2\rangle\langle a_2|$ (83% belief to buy, 17% belief to sell)
- $S_3 = \left( ((I * Y) + H) + I \right) \rightarrow \hat{V} = 0.97|a_1\rangle\langle a_1| + 0.03|a_2\rangle\langle a_2|$ (97% belief to buy, 3% belief to sell)
- $S_4 = \left( ((I * Y) + H) + Y \right) \rightarrow \hat{V} = 0.5|a_1\rangle\langle a_1| + 0.5|a_2\rangle\langle a_2|$ (50% belief to buy, 50% belief to sell)
- $S_5 = \left( D * (Y + (X + T)) \right) \rightarrow \hat{V} = 0.43|a_1\rangle\langle a_1| + 0.57|a_2\rangle\langle a_2|$ (43% belief to buy, 57% belief to sell)
- $S_6 = \left( D * (Y + (I + T)) \right) \rightarrow \hat{V} = 0.68\langle a_1| + 0.32|a_2\rangle\langle a_2|$ (68% belief to buy, 32% belief to sell)
- $S_7 = \left( D * \left( Y + \left( (D * (S + Z)) + T \right) \right) \right) \rightarrow \hat{V} = 0.16|a_1\rangle\langle a_1| + 0.84|a_2\rangle\langle a_2|$ (16% belief to buy, 84% belief to sell)
- $S_8 = \left( D * \left( Y + ((T + H) + T) \right) \right) \rightarrow \hat{V} = 0.24|a_1\rangle\langle a_1| + 0.76|a_2\rangle\langle a_2|$ (24% belief to buy, 76% belief to sell)

Based on the degrees of beliefs they can be further grouped into 4 groups:

- Group One: the trader is close to evenly split (50/50) of buy and sell, which includes strategies $S_4$ and $S_5$
- Group Two: the trader is heavily committed to buying, which includes strategies $S_1$, $S_2$, and $S_3$
- Group Three: the trader is heavily committed to selling, which includes strategies $S_7$, and $S_8$
- Group Four: the trader is leaning towards buying but is not that fully committed yet, in this case is only strategy $S_6$

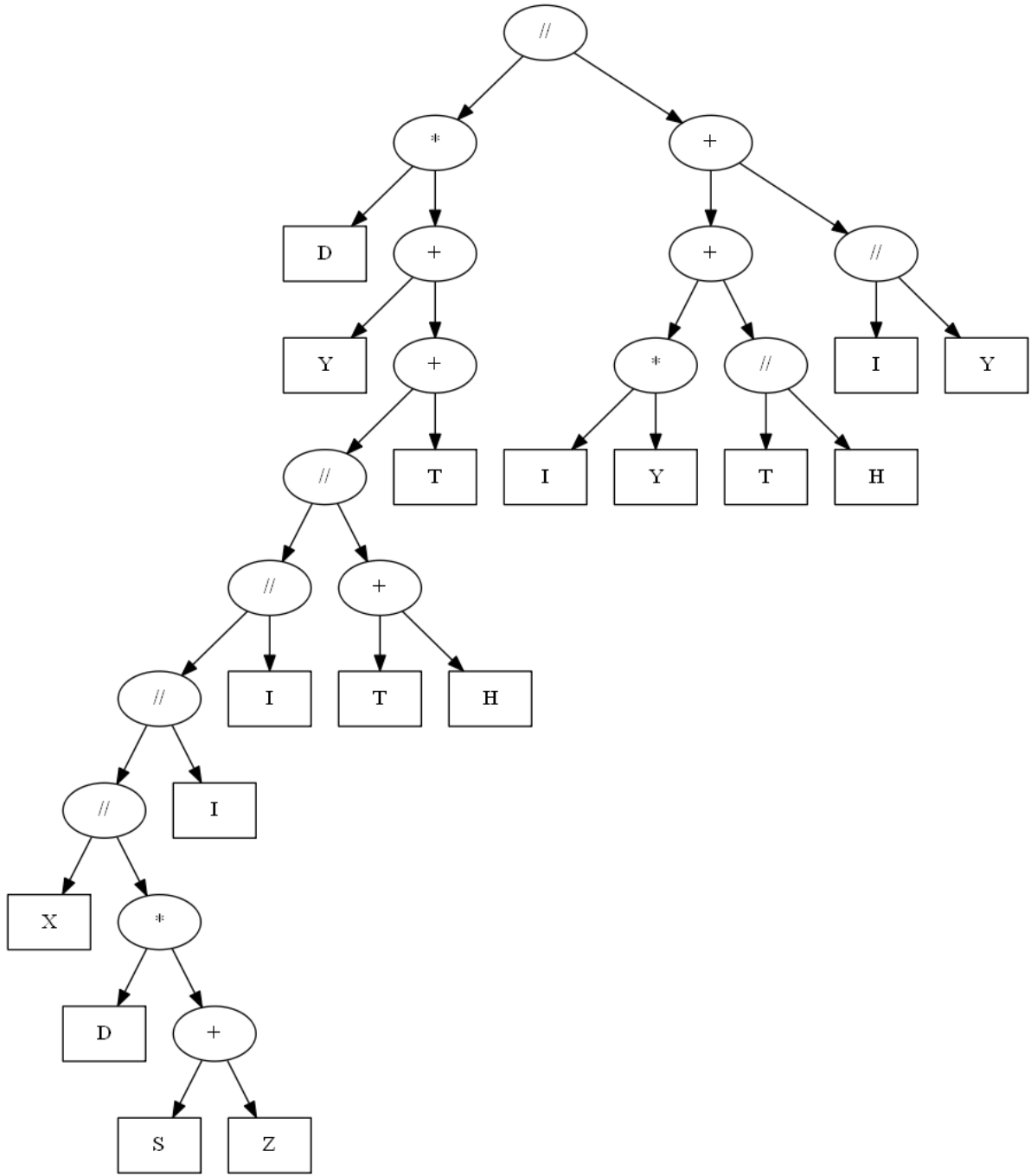

**Figure 4** An optimized qDT

This qDT has a mixed strategy comprising of the eight strategies to choose from. If a strategy from group one is chosen, with its close to 50/50 probability of either buy or sell is selected, then in this case no valuable information will be obtained because it's totally uncertain; or if one of the strategies from the group two is selected with almost 100% surety to buy, then maximum information is obtained; or if a strategy from group three is selected with close to 100% surety to sell, then maximum information is also obtained; and since strategy $S_6$ is the lone strategy of not fully committed to a decision, though leaning towards one, the best information that it can provide is to just follow your instinct to choose. One thing that is clear from these groups of strategies is that, if the belief of buying or selling is high then it is best to follow through on that belief. If it is hovering between 50/50 it's best to just wait (don't do anything), or just flip a coin to decide. It's just like how the old sayings go, "trust your gut", and "when in doubt, hit the luck." qDT can only pick one of the eight strategies to actually execute. It can't pick two, three, or all eight strategies at once, it has to choose one and one only.

Once it selects that one strategy, it is still with a certain degree of belief of choosing whether to buy or sell with the chosen strategy. For example, if it chooses strategy $S_1$, then there's an 88% degree of belief to buy, and a 12% degree of belief to sell.

| Market State | Strategy | Action | Degree of Belief | Gain or Loss |
|---|---|---|---|---|
| down | $S_1$ | buy | 88% | -8.0 |
| up | $S_1$ | buy | 88% | 10.0 |
| up | $S_8$ | buy | 24% | 2.0 |
| up | $S_3$ | buy | 97% | 3.0 |
| up | $S_5$ | sell | 57% | -5.0 |
| up | $S_8$ | buy | 24% | 3.0 |
| up | $S_1$ | buy | 88% | 16.0 |
| down | $S_7$ | sell | 84% | 2.0 |

**Table 3** Details of the first eight transactions.

The table 3 illustrates the first 8 transactions our qDT made and which strategies it took. In the table, in the first trade that it completed, strategy $S_1$ was chosen, with an 88% degree of belief to buy, but since the state of the market was trending downwards at the time of the transaction, it resulting in a deficit of -8. And in the second trade, strategy $S_1$ was also chosen, but this time the market was trending upwards, which resulted in a profit of 10. Trade number 3 saw our qDT take strategy $S_8$, with only a 24% degree of belief to buy, and though the market was trending up at the time, there was a gain of 2. Following this logic, the rest of the trades are pretty much self-explanatory, with the qDT ending off with five "victories" out of the first eight transactions, as illustrated by the full table above. For the total 218 trades that the qDT made, the odds are around 60% in favor, which is up from the expected 50/50 chance of winning or losing in the market.

## Discussion

$$\text{market} = \{q_i, x_i, i = 1, 2, \cdots N\} \tag{16}$$

The $q_i$ depicted is the dynamic state of the market, basically if it's up or down at any given time, but at set intervals, i.e., 1 day, 2 days, 3 days. And the $x_i$ that is depicted above the closing price that is determined by all the traders participating in the market, their trading, whether it be buy or sell, will determine the sequence of the price, {x, i=1, 2,…,N}, which represents the "path" of the market, where we see is it as the market's volatility.

Thus, the traditional way is to use stochastic differential equations to describe the market's volatility and treat it as a stochastic process. Economists have used this "classical" way of formulating a mathematical model of the market and how traders make decisions under uncertainty. But this stochastic math model of the market comes short in some ways, the most notably being that it can't really describe what actually happens in the real world because of all the uncertainty at play. Everyone makes a decision with at least some influence of the past, which means to some degree, we all remember and factor in past decisions when making future ones, in contrast to the Markov decision process where all decision-makers have amnesia.

The constant problem that decision-makers' face is the challenge of making decisions under the influence of an dynamically-changing market, in which it can be seen as a game between trader(s) and the market, a game of question and answer, with the market "asking" the questions, in which the trader responds (answers) through their actions. This can be described as:

The "questions" posed by the market: $\{q_i, x_i\}, i = 1, 2, \cdots N$

To answer the market's "questions", the trader uses yes/no logic in quantum decision trees: $\{qDT_j, j = 1, 2, \cdots M\}$

We can look at the game between trader and market in this way: there is a sequence of "choices" made that are posed as "questions" by market, and the traders' must "answer" those "questions" by selecting a sequence of actions guided by optimized strategies to "beat" the market. The mathematical way to explain the real world is not very ideal, especially in the context of a "game" between trader and market. But what is, and may be the ideal solution to explain this Q&A "game" is Darwin's theory of evolution. Whenever we make decisions, we do so to maximize our expected value to the greatest extent. The reason why we maximize our expected value is the equivalent to survival. And we can do this by proposing a hypothetical robot "trader" that will take place in playing this game of art and strategy with the market. The robot "trader" will learn the historical data that was posed in question form by the market. By utilizing the yes/no logic of quantum decision tree, then applying it to evolve the natural dynamic rules to formulate satisfying strategies to guide robot trader in the "game" with the market. The description of utilizing a computer simulation process to answer the market's question, instead of the traditional stochastic differential equations approach is as follows:

$$\text{historical data (market)} \xleftarrow{\text{learn}} \text{computational simulation (robot "trader")} \begin{cases} \xrightarrow{\text{evolve}} \text{market dynamics rules} \\ \xrightarrow{\text{reconstruct}} \text{market price "path"} \end{cases}$$

The observed data of the market (the questions posed by nature): $\{(q_1, x_1), (q_2, x_2), ..., (q_n, x_n)\}$, in which $q_n$ denotes the state of the market (up or down), and $x_n$ denotes the market's price. Then we get to the robot trader's learning and optimization process, in which it "computes" a sequence of data that will be "matched" with the market's state in order to see if it "answered" correctly: $\{(a_1, y_1), (a_2, y_2), ..., (a_n, y_n)\}$, in which $a_n$ denotes the action (buy or sell) taken by the robot, and $y_n$ is the final price that is calculated based on the action, $a_n$, that the robot "trader" took.

This process works like this: utilizing the qDT that produces a group of strategies to choose from, each one of those strategies presents two actions to take, which in turn each have a probability degree of subjective belief, and the robot "trader" can only choose one of the actions of one of the strategies, and one only. It is then with that, then the price that the trader will buy or sell at is computed that is denoted by $y_n$. Throughout the process of learning, the robot "trader" is able to compute a price that will be consistent with the market price, which will mean that the $(q_n, x_n)$ and the $(a_n, y_n)$ will "match" each other, then the trader will have "answered" the "questions" posed by nature.

The absolute deviation function is defined as:

$$D_i = |y_i - x_i| \tag{17}$$

Where $x_i$ is the market actual price, $y_i$ is the computed price of the market by the robot "trader". We take the absolute deviation as negative feedback to fine tune the fitness function as below:

$$f_{fitness} = \sum_{i=1}^{N} qEV_i/D_i \tag{18}$$

By maximizing $qEV_i$ (quantum expected value) to "guess" the current dynamic state of the market, in which quantum decision tree (qDT) is utilized to assess the market's future state, and $D_i$ (negative feedback), which is utilized to adjust the "calculated market price" based on assessing the trader's past experience to match the market's actual price.

The Table 4 is the 218 trades, in which the market state of up (0) or down (1) is presented as historical data to the robot "trader" to learn from, labeled in the second column. Then using the fitness function that was just defined above, the robot "trader" is "trained" by 25 times that produces 25 quantum decision trees (qDT), labeled as $qDT_1 \sim qDT_{25}$. Each of those qDTs then becomes the trader's "experience" regarding the market through learning the market's historical data. Altogether, the 25 qDTs are the trader's entire experience (knowledge) about the market.

| trade | market state | qDT$_1$ | ... | qDT$_{18}$ | ... | qDT$_{25}$ |
|---|---|---|---|---|---|---|
| 1 | 1 | 0 | ... | 0 | ... | 1 |
| ⋮ | ⋮ | ⋮ | ⋮ | ⋮ | | ⋮ |
| 100 | 0 | 0 | ... | 0 | ... | 0 |
| ⋮ | ⋮ | ⋮ | ⋮ | ⋮ | | ⋮ |
| N=218 | 0 | 0 | ... | 0 | ... | 0 |

**Table 4** Details of the trades made by robot.

Over the course of the 218 trades, each qDT's chance of winning is around 60%, which equates to approximately 130 times win. Just like how we rely on our entire past experiences to make decisions, our robot also needs to rely on all of its past experiences, which is the experience it gained from all of the qDTs as a whole. According to Table 4, by applying majority rules, we can let our robot trader determine the market's state based on the experience obtained from 25 qDTs. The rules are as follows:

1. If more than half of the qDTs decide that the market is up, then the robot trader believes the market is up.

2. If more than half of the qDTs decide that the market is down, then the robot trader believes the market is down.

By applying majority rules, the robot trader now has its chance of winning increased to about 90%, equating to around 195 wins. This is reflected in the graph below.

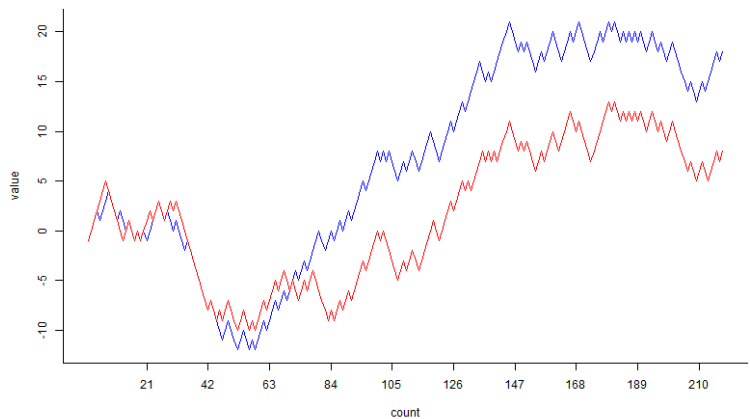

**Figure 5** The "trajectories" of market reconstructed by 25 qDTs

In the graph above the real market's data curve is the red line, and the robot trader's trades is the blue line. And as we can see, the robot trader's trades were almost in complete sync with the actual market. So, this shows that our robot trader seems to have "decoded" the market's "enigma" through the learning of historical data, training by the evolution algorithm, and executing its actions with the quantum decision trees. If the market acts the exact same way according to its historical data in the future, then our robot trader can definitely continue to beat the market, but in reality, the market can have an infinite number of paths it can take, and if what path it takes is completely random, it is still very hard, if not impossible to predict the future of the market. There are infinite possible paths for us to take, but we can only certainly live to see what will happen in only one of them because we cannot have prior information of the market.

## Data availability
The data that support the findings of this study are available on request.

## Code availability
Codes are available upon request.

## Author contributions statement

All authors conducted the research and contributed to the development of the model. HX contributed as an expert in quantum theory and non-linear science. LX contributed to the research from the aspects of machine learning, decision theory and wrote the code. KX wrote this manuscript and did data analysis. All authors reviewed the manuscript.

## Additional information

**Competing interests**: The authors declare no competing interests.